\documentclass[aps,prl,10pt,amsmath,twocolumn,floatfix,superscriptaddress]{revtex4-1}
\usepackage[dvipsnames]{xcolor}
\usepackage[colorlinks=true,bookmarks=false,linkcolor=NavyBlue,urlcolor=NavyBlue,citecolor=NavyBlue,breaklinks]{hyperref}
\usepackage{amsmath}
\usepackage{amsfonts}
\usepackage{graphicx}
\usepackage{times}
\usepackage{mathptmx}
\usepackage{braket}

\newcommand{\real}{\operatorname{Re}}

\newcommand{\parti}[2]{\frac{\partial #1}{\partial #2}}

\newcommand{\intall}{\int_{-\infty}^{\infty}}

\newcommand{\bk}[1]{\left(#1\right)}

\newcommand{\trace}{\operatorname{tr}}

\newcommand{\expect}{\mathbb E}

\begin{document}

\title{Semiclassical Theory of Superresolution for Two Incoherent
Optical Point Sources}

\author{Mankei Tsang}
\email{mankei@nus.edu.sg}
\affiliation{Department of Electrical and Computer Engineering,
  National University of Singapore, 4 Engineering Drive 3, Singapore
  117583}

\affiliation{Department of Physics, National University of Singapore,
  2 Science Drive 3, Singapore 117551}

\author{Ranjith Nair}
\affiliation{Department of Electrical and Computer Engineering,
  National University of Singapore, 4 Engineering Drive 3, Singapore
  117583}

\author{Xiao-Ming Lu}
\affiliation{Department of Electrical and Computer Engineering,
  National University of Singapore, 4 Engineering Drive 3, Singapore
  117583}

\date{\today}


\begin{abstract}
  Using a semiclassical model of photodetection with Poissonian noise
  and insights from quantum metrology, we prove that linear optics and
  photon counting can optimally estimate the separation between two
  incoherent point sources without regard to Rayleigh's criterion.
  The model is applicable to weak thermal or fluorescent sources as
  well as lasers.
\end{abstract}

\maketitle
Lord Rayleigh suggested in 1879 that two incoherent optical point
sources should be separated by a diffraction-limited spot size for
them to be resolved \cite{rayleigh}. This criterion has since become
the most influential measure of imaging resolution. Under the modern
advent of rigorous statistics and image processing, Rayleigh's
criterion remains a curse.  When the image is noisy, necessarily so
owing to the quantum nature of light \cite{mandel}, and Rayleigh's
criterion is violated, it becomes much more difficult to estimate the
separation accurately by conventional imaging methods
\cite{bettens,vanaert,ram}. Modern superresolution techniques in
microscopy \cite{hell,betzig,moerner} can circumvent Rayleigh's
criterion by making sources radiate in isolation, but such techniques
require careful control of the fluorescent emissions, making them
difficult to use for microscopy and irrelevant to astronomy.

Here we show that, contrary to conventional wisdom, the separation
between two incoherent optical sources can be estimated accurately via
linear optics and photon counting (LOPC) even if Rayleigh's criterion
is severely violated. Our theoretical model here is based on the
semiclassical theory of photodetection with Poissonian noise, which is
a widely accepted statistical model for lasers \cite{mandel}
as well as weak thermal \cite{goodman_stat,labeyrie} or fluorescent
\cite{pawley,ram} light in astronomy and microscopy. The semiclassical
model is consistent with the quantum model proposed in Ref.~\cite{tnl}
for weak incoherent sources and the mathematical formalisms are
similar, but the semiclassical model has the advantage of being
applicable also to lasers, which are important sources for
remote-sensing, testing, and proof-of-concept experiments. The
semiclassical theory also avoids a quantum description of light and
offers a more pedagogical perspective. Compared with the full
semiclassical theory in Ref.~\cite{sliver}, the Poissonian model is
invalid for strong thermal sources but more analytically
tractable.

Consider $J$ optical modes and a column vector of complex field
amplitudes $\alpha = (\alpha_1,\dots,\alpha_J)^\top$ within one
coherence time interval. The amplitudes are normalized such that
$|\alpha_j|^2$ is equal to the energy in each mode in units of quanta.
The central quantity in statistical optics is the mutual coherence
matrix \cite{mandel,goodman_stat}
\begin{align}
\Gamma &\equiv \expect\bk{\alpha\alpha^\dagger}, 
\end{align}
where $\dagger$ denotes the complex transpose and $\expect$ denotes
the statistical expectation. We also define 
$\epsilon \equiv \expect\bk{\alpha^\dagger\alpha}  = \trace\Gamma$
as the mean total energy, $\trace$ as the trace, and
\begin{align}
g &\equiv \frac{\Gamma}{\trace\Gamma}
\end{align}
as the correlation matrix. $g$ is positive-semidefinite with unit trace
and typically called $g^{(1)}$ in statistical optics.

Suppose that we process the optical fields with lossless passive
linear optics, the input-output relations of which are characterized
by a unitary scattering matrix $F$. The output mutual coherence matrix
becomes $F\Gamma F^\dagger$ \cite{mandel,goodman_stat}. The average
energy in the $j$th output optical mode accumulated over $M$ coherence
intervals is the $j$th diagonal component of $F\Gamma F^\dagger$ times
$M$, which can be written as
\begin{align}
\bar n_j &= M e_j^{\dagger} F\Gamma F^\dagger e_j = N p_j,
&
N &\equiv M\epsilon, &
 p_j &\equiv \trace \Pi_j g,
\label{pj}
\end{align}
where $e_j$ is a column unit vector with $e_{jk} = \delta_{jk}$, $N$
is the average total energy, $p_j$ is a normalized output energy
distribution with $\sum_j p_j = 1$, and
$\Pi_j = F^\dagger e_j e_j^\dagger F$ is a projection measure with the
completeness property $\sum_j \Pi_j = I$, $I$ being the identity
matrix.

Consider photodetection at the output modes. Assume that the
probability distribution of the $ n\equiv (n_1,\dots,n_J)^\top$
photoelectrons is Poissonian:
\begin{align}
P(n) &= \prod_j \exp(-\bar n_j)\frac{\bar n_j^{n_j}}{n_j!}.
\label{poisson}
\end{align}
This is the standard shot-noise model for weak thermal
\cite{goodman_stat,labeyrie} or fluorescent \cite{pawley,ram} sources
with $\epsilon \ll 1$, in which case it is also consistent with the
quantum model in Ref.~\cite{tnl}. Bunching or antibunching effects
would lead to slightly non-Poissonian statistics, but they are
negligible for typical sources
\cite{mandel,goodman_stat,labeyrie}. Beyond weak sources, the
Poissonian model is also applicable to ideal lasers with arbitrary
$\epsilon$ \cite{mandel}. This is convenient not only for laser
sensing applications, but also for testing and proof-of-concept
purposes in telescopy and microscopy.

Suppose that $g$ depends on a vector of unknown parameters
$\theta \equiv (\theta_1,\dots,\theta_R)^\top$.  Given a measurement
record $n$, define the estimator vector as $\check\theta(n)$ and the
mean-square-error matrix as
$\Sigma \equiv
\expect\bk{\check\theta-\theta}\bk{\check\theta-\theta}^\top$, where
the expectation is with respect to $P(n)$. The Cram\'er-Rao
bound is given by the matrix inequality \cite{vantrees}
\begin{align}
\Sigma &\ge \mathcal J^{-1},
&
\mathcal J_{\mu\nu} &\equiv \sum_n \frac{1}{P(n)}\parti{P(n)}{\theta_\mu}
\parti{P(n)}{\theta_\nu},
\end{align}
where $\mathcal J$ is the Fisher information matrix. For the
Poissonian model given by Eq.~(\ref{poisson}), the matrix can be
significantly simplified as
\begin{align}
\mathcal J_{\mu\nu} &= \sum_j \frac{1}{\bar n_j}\parti{\bar n_j}{\theta_\mu}
\parti{\bar n_j}{\theta_\nu} = 
N \sum_j \frac{1}{p_j}\parti{p_j}{\theta_\mu}\parti{p_j}{\theta_\nu}.
\label{Jpoisson}
\end{align}
Notice that Eq.~(\ref{Jpoisson}) has the same expression as the Fisher
information with respect to a probability distribution. Since $p_j$
can be expressed in terms of a projection measure $\Pi_j$ and a
unit-trace positive-semidefinite matrix $g$ according to
Eq.~(\ref{pj}), we can borrow the mathematical formalism from quantum
metrology \cite{helstrom,hayashi05} to write immediately
\begin{align}
\mathcal J &\le \mathcal K(g),
\\
\mathcal K_{\mu\nu} &\equiv N \real \trace \mathcal L_\mu \mathcal L_\nu g,
\quad
\parti{g}{\theta_\mu} = \frac{1}{2}\bk{\mathcal L_\mu g + g \mathcal L_\mu},
\end{align}
where $\mathcal K(g)$ is the Helstrom-Fisher information matrix in
terms of $g$. This upper bound quantifies the maximum information that
can be extracted from the light source via any LOPC with Poissonian
noise. A connection with the quantum model in Ref.~\cite{tnl} can be
made by observing that the one-photon density matrix in the quantum
model is approximately $g$ under the $\epsilon \ll 1$ assumption,
although the quantum description and the $\epsilon \ll 1$ assumption
are unnecessary here, as long as the Poissonian model holds. This
demonstrates the power of quantum metrology for an essentially
classical problem.

The rest of the theory is almost the same as that in Ref.~\cite{tnl}
mathematically, with identical physical conclusions.  Taking the
continuous-space limit for a one-dimensional image plane, $g$ becomes
the correlation function $g(x,x')$ with normalization
$\intall dx g(x,x) = 1$, and the intensity distribution for direct
imaging is $N g(x,x)$.  For a diffraction-limited point-spread
function $\psi(x)$ and two point sources at $X_1$ and $X_2$ with
random relative phase,
$g(x,x') = [\psi(x-X_1)\psi^*(x'-X_1)+\psi(x-X_2)\psi^*(x'-X_2)]/2$.
In terms of the separation parameter $\theta_2 = X_2 - X_1$,
$\mathcal J_{22}^{(\rm direct)}$ for direct imaging suffers from
Rayleigh's curse and approaches zero for $\theta_2 \to 0$
\cite{bettens,vanaert,ram}. Meanwhile, the Helstrom-Fisher information
$\mathcal K_{22}$ is constant for a real $\psi(x)$ and given by
$\mathcal K_{22} = N\Delta k^2$, with $\Delta k^2$ being the
momentum-space variance of $\psi(x)$.

\begin{figure}[htbp!]
\centerline{\includegraphics[width=0.42\textwidth]{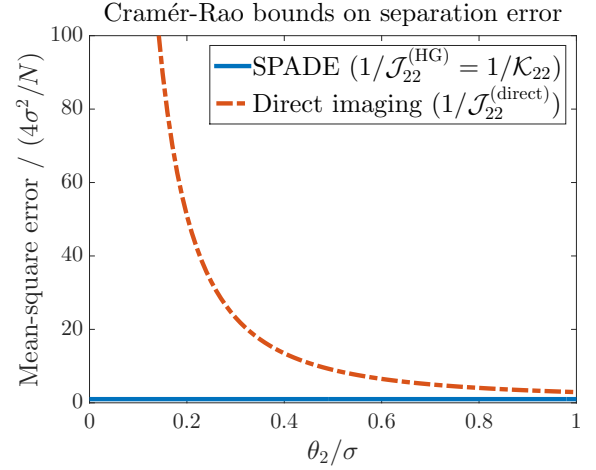}}
\caption{\label{crb}Cram\'er-Rao bounds for SPADE and direct imaging
  with a Gaussian point-spread function.}
\end{figure}

Assuming a Gaussian $\psi(x)$ with width $\sigma = 1/(2\Delta k)$,
spatial-mode demultiplexing (SPADE) of the image-plane fields in the
Hermite-Gaussian modes leads to
$\mathcal J_{22}^{(\rm HG)} = N/(4\sigma^2) = \mathcal K_{22}$, which
overcomes Rayleigh's curse and attains the maximal
information. Figure~\ref{crb} compares the Cram\'er-Rao bounds for
SPADE and direct imaging, demonstrating the substantial improvements
deliverable by SPADE. The other discussions and simulations in
Ref.~\cite{tnl} concerning binary SPADE, maximum-likelihood
estimation, and misalignment remain valid here. The formalism is also
applicable to the SLIVER scheme in Ref.~\cite{sliver}, meaning that
the scheme must also work for incoherent laser sources, and
generalizable to any parameter estimation problem.

This work is supported by the Singapore National Research Foundation
under NRF Grant No.~NRF-NRFF2011-07.

\bibliography{research}
\end{document}